\documentclass[prl,twocolumn]{revtex4}
\usepackage{epsfig,graphicx}
\usepackage[dvipdfm,breaklinks=true,colorlinks=false,citecolor=black,urlcolor=blue]{hyperref}
\begin{document}

\title{Quantum dot as thermal rectifier}

\author{R. Scheibner, M. K{\"o}nig, D. Reuter$^*$, A.D.
Wieck$^*$, H. Buhmann, and L.W. Molenkamp}

\affiliation{Physikalisches Institut (EP3), Universit\"at
W\"urzburg, Am Hubland, 97074 W\"urzburg, Germany}

\affiliation{$^*$Lehrstuhl f{\"u}r Angewandte Festk{\"o}rperphysik,
Ruhr-Universit{\"a}t Bochum, Universit{\"a}tsstra{\ss}e 150, 44780 Bochum,
Germany}

\date{\today}

\begin{abstract}
We report the observation of thermal rectification in a semiconductor quantum dot, as inferred from the asymmetric line shape of the thermopower oscillations. The asymmetry is observed at high in-plane magnetic fields and caused by the presence of a high orbital momentum state in the dot.
\\


\end{abstract}

\maketitle


Recently, the field of thermoelectricity and solid state thermionics has gained renewed attention, due to advances in growth and fabrication of complex compounds, mesoscopic devices, and nanostructures \cite{Giazotto:RMP:78:217:2006,ODwyer:PRL:72:205330:2005}. The main idea is to enhance the efficiency of macroscopic devices by the control of the energy transport on a microscopic scale. For example, a recent proposal to build a thermal rectifier in a non-linear lattice attracted much attention \cite{Terraneo:PRL:88:094302:2002}. An actual nanoscale solid state thermal rectifier using tailored carbon and boron nitride nanotubes was recently demonstrated \cite{Chang:Science:314:1121:2006}. Here, we present yet another way to obtain thermal rectification, utilizing the thermoelectric properties of a semiconductor quantum dot (QD).

QDs are the smallest possible thermoelectric devices. In the Coulomb blockade (CB) regime, their transport properties are highly nonlinear and strong thermoelectric signals result \cite{Staring:EPL:22:57:1993}. The thermopower $S_{\rm QD}$ parametrizes the electric response $V_{\rm T}$ of a QD to an applied temperature difference $\Delta T$ ($V_{\rm T}= - S_{\rm QD}\Delta T$). It can be related to the average energy $\langle E \rangle$ of the carriers by $S_{\rm QD} =- \langle E \rangle/eT$. 

Transport through a QD depends crucially on the coupling strength of the dot states to the leads, i.e.\ the wave function overlap of the localized and free states. While asymmetries in this coupling have been observed in electrical transport measurements (see for example  \cite{Weis:PRL:71:4019:1993,Schleser:PRB:72:035312:2005}), their
influence on the thermoelectric properties still is outstanding. 
In this paper we compare thermovoltage and conductance measurements on a gate-defined QD in the Coulomb blockade (CB) regime. In high magnetic fields, applied in the plane of the two dimensional electron gas (2DEG), a suppression of carrier transport is observed for certain single electron tunneling (SET) resonances. The corresponding thermoelectric signal exhibits a strong asymmetry. An analysis of this asymmetry reveals that the QD acts as a thermoelectric rectifier.


\begin{figure}
\centering
\includegraphics[width = 8.6 cm]{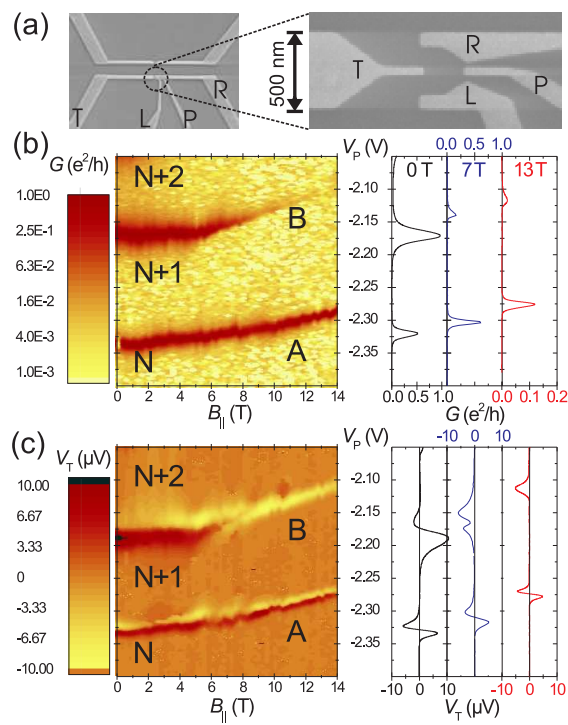}
\caption{(a) SEM image of the QD structure; enlarged QD region on the right. Schottky-gates are labeled T, R, L, and P. (b) Logarithmic grey scale plot (color online) of the conductance $G$ as a function of $B_\parallel$ and $V_{\rm P}$. (c) Corresponding grey scale plot of the thermovoltage (linear scale). Traces at 0, 7 and 13 T are given separately.}\label{fig1}
\end{figure}

Figure~\ref{fig1}~(a) shows a scanning electron microscope image
of the QD structure. This structure is fabricated by electron-beam
lithography on a GaAs/(Al,Ga)As heterostructure with a 2DEG carrier density of $n_s = 2.3\times 10^{15}$~m$^{-2}$ and a nominal mobility of $\mu=100$~m$^2$/(Vs). The QD is embedded at the border of a 2 $\mu$m wide and 20 $\mu$m long electron heating channel \cite{Scheibner}. The lithographic diameter of the QD is approximately $250$~nm and the number of electrons can be tuned between 20 and 30 as a function of applied plunger gate-voltage, $V_{\rm P}$. Low-frequency $(\nu=13\text{ Hz})$ lock-in techniques are used for electrical and thermoelectrical measurements. A current heating technique is used to create a temperature difference of $\Delta T \approx 30$~mK  across the QD. The temperature difference and the electron gas temperature ($T_e= 80$~mK) are determined independently by analyzing SET resonances as well as the temperature and heating current dependence of the universal conductance fluctuations of the heating channel.


Figure~\ref{fig1}~(b) presents a grey (color online) scale plot of the zero bias conductance in the regime of two SET resonances at $V_{\rm P} = -2.33$~V (resonance A) and $V_{\rm P} = - 2.17$~V (resonance B) as a function of the plunger gate voltage ($V_{\rm P}$) and magnetic field ($B_{\parallel}$) applied parallel to the 2DEG plane. Resonance A is typical for standard SET CB behavior, whereas the new physics this paper focuses on is inferred from the behavior of resonance B. The resonances occur where the chemical potentials of source and drain contact leads align with that of the QD, i.e., $\mu_{\rm S}=\mu_{\rm D}=\mu_{n+1;n}=E_{n+1,0}-E_{n,0}$, where $E_{n+1,0}$ and $E_{n,0}$ denote the ground state (0) energy of $n+1$  and $n$ electrons on the QD, respectively. Both SET peaks shift to less negative gate voltages with increasing magnetic fields. This diamagnetic shift is due to a different magnetic field dependence of the energy states in the QD and in the leads \cite{Weis:PRL:71:4019:1993, Duncan:APL:77:2183:2000}. More striking is the observation that the amplitude of resonance B decreases strongly with increasing magnetic field and almost vanishes for $B_{||} > 8$~T while the amplitude of resonance A remains almost constant. For clarity, single traces of $G(V_{\rm P})$ at $B_\parallel = 0$, 7, and 13~T are shown in the right panel of Fig.~\ref{fig1}~(b).

The corresponding thermovoltage is shown in Figure~\ref{fig1}~(c). Dark and bright regions correspond to large positive and large negative thermovoltage signals, respectively. SET resonances occur at the sharp transition from positive to negative thermovoltages.  The observed diamagnetic shift is clearly visible in the thermovoltage measurement. While the thermovoltage signal of resonance A shows a uniform behavior for the whole magnetic field range, the thermovoltage of resonance B exhibits some significant changes for $B\ge 5$~T. For this resonance, the characteristic positive-to-negative $V_{\rm T}$ transition turns into a double peak structure with negative amplitudes for $5\text{ T}<B<8$~T. For $B>8$~T, only a single negative thermovoltage signal remains. In contrast to what we observe for this peak in the conductance measurements, the amplitude of the negative thermovoltage signal remains approximately unchanged [right panel of Fig.~\ref{fig1}~(c)]. In general, such an asymmetric thermovoltage indicates that the electron-hole symmetry of a SET process in a QD is broken \cite{Beenakker:PRB:46:9667:92,Turek:PRB:65:115332:2002}.

\begin{figure}
\centering
\includegraphics[width = 8.6 cm]{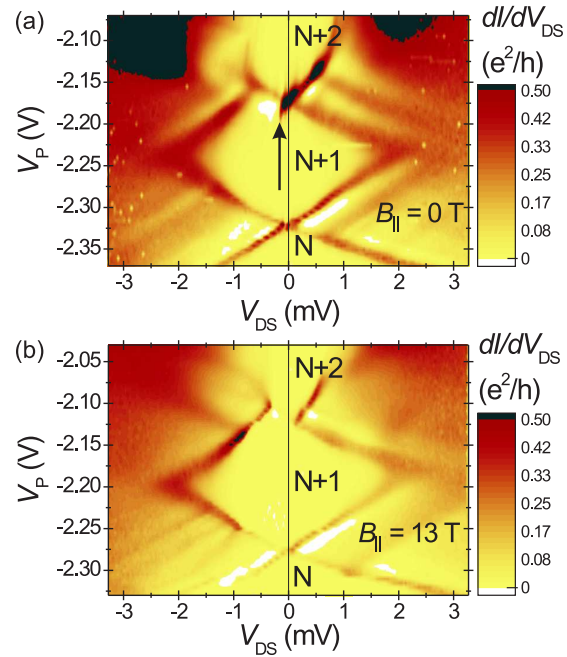}
\caption{(color online) Grey scale plot of the differential conductance as a function of $V_{\rm DS}$ and $V_{\rm P}$ at (a) $B_\parallel=0$~T and (b) $B_\parallel=13$~T. The arrow marks the transition from a suppressed to a high differential conductance.}\label{fig2}
\end{figure}

The broken symmetry is even more obvious in the electron addition spectrum, $G(V_{\rm DS},V_{\rm P})$, which is displayed in Fig.~\ref{fig2} for $B = 0$~T and $B=13$~T. In the diamond shaped  regions, the dot has a constant electron number ($\dots N, N+1, N+2, \dots$). The dark lines (non-zero differential conductance) delimiting the diamonds are due to the SET resonances, whose positions depend on source-drain and plunger-gate voltage. 
Resonance lines outside the CB diamonds originate from transport through excited states ($\mu_{n+1,i;n,j}$), where $(i,j)\, \epsilon \, \mathbf{N}^+$ \cite{Foxman:PRB:47:10020:1993}. The appearance of areas with negative differential conductance (white areas) is typically a signature of the occurrence of blocking mechanisms of various kinds \cite{Weinmann:PRL:74:984:1995,Destefani:PRB:65:235314:2002}. Especially for $B=0$, the SET resonance $\mu_{(N+2;N+1)}$ exhibits a negative differential conductance for  $V_{\rm DS} < 0$~V and $V_{\rm P}< -2.21$~V [onset indicated by an arrow Fig.~\ref{fig2}~(a)]. The resonance line with a positive slope involving the ground state ($\mu_{N+2,0;N+1,0}$), is absent. This blocking of the electrical transport for negative $V_{DS}$ indicates that the drain contact is aligned with a QD state which couples asymmetrically with the contacts. At $B_\parallel = 13$~T [Fig.~\ref{fig2}~(b)], an asymmetric transport gap opens for the zero bias $N+2\leftrightarrow N+1$ transition, i.e., a new asymmetrically coupled state becomes the QD ground state $\mu^\ast_{(N+2;N+1)}$, which strongly suppresses the charge carrier transport, especially for negative bias voltage. Since in our experiments, source and drain contacts are adjusted to equal transmission, the observed transport asymmetry has to be related to the intrinsic symmetry of the QD states.

%
%

The experimental observation can be explained \cite{Weis:PRL:71:4019:1993,Schleser:PRB:72:035312:2005} by considering two nearly degenerate QD states. The $B=0$ ground state, $\mu_{(N+2;N+1)}$, couples symmetrically to source and drain contacts while, the second state, $\mu_{(N+2;N+1)}^\ast$, couples more strongly to the drain than to the source. This situation is schematically depicted in Fig.\ 4 c). For negative $V_{DS}$ the drain contact will populate both states and thus the low lying symmetric state will become blocked while for positive $V_{DS}$, electrons that originate from the source will only populate the symmetric ground state and transport will be possible.

In a magnetic field parallel to the plane of the 2DEG, the Zeeman effect introduces an energy shift to the QD states according to $E_{Z}= m_J g\mu_BB$, where $m_J$ is the magnetic quantum number of the QD state with total angular momentum $J=L+S$, $g$ the associated g-factor, and $\mu_B$ the Bohr magneton. In an in-plane magnetic field the orbital component of the electronic wave function remains almost unchanged, which ensures that the coupling to the leads of the various states is not affected. However, the field may very well affect the energetic ordering of the orbital states. From the design of our QD one infers that a symmetric coupling to the leads is more likely for a symmetric QD state with orbital angular momentum $L=0$, and that an asymmetric coupling must involve states with $L \ne 0$. In combination with the above observations this implies that, in order to explain the experiments, a magnetic field must change the energetic ordering of these states.

\begin{figure}
\centering
\includegraphics[width = 8.6 cm]{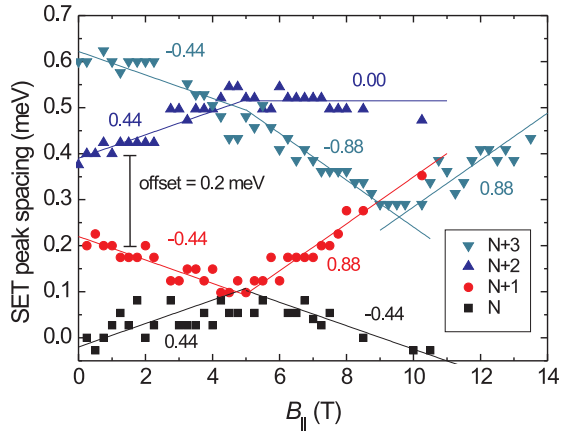}
\caption{(color online) SET peak spacings as a function of $B_\parallel$ for constant electron numbers. The curves are
distributed equally spaced ($\Delta = 0.2 meV$) on a scale between 0 and 0.6~meV with increasing $N$. The lines are linear fits with slopes of indicated multiples of $\mu_B$.}\label{fig3}
\end{figure}

Information about the energetic ordering of the QD states can be obtained by analyzing the magnetic field dependent energy separation of the SET conductance peaks \cite{Duncan:APL:77:2183:2000,Tarucha:PRL:77:3613:1996}. In Figure~\ref{fig3} the energy spacings between neighboring SET resonances is plotted as a function of $B_{\parallel}$. The $N$ and $N+1$ states refer to the states with the same index in Figs.\ 1 and 2. For clarity, the curves are distributed equally spaced ($\Delta = 0.2$ meV) on a scale between 0 and 0.6~meV with increasing $N$. The curves exhibit a piecewise linear behavior, with slopes that can be expressed in integer multiples of $g_{GaAs}=(-0.44)$, i.e., the g-factor of the QD states. 

The slope of the SET peak spacings is equivalent to the difference in Zeeman energy, $\Delta E_Z = \Delta m_J g\mu_B B$, for the two involved QD states. For example, a slope of $+0.44\mu_B$ is obtained for $\Delta m_J = m_{J(n+1)}-m_{J(n)}=1$. A change of the slope indicates a rearrangement of the spin configuration and a state with a different $m_{J^*(n+1)}$ becomes the new ground state. Such rearrangements are observed at $B = 5$ and $9.5$ T. Of particular interest is the increase of the slope for the $N+1$ state to $0.88\, \mu_B$ at $B = 5$ T which indicates that indeed a QD state with $L \ne 0$ is involved in this transitions.\\ 


In order to verify that the simple picture of an asymmetrically coupled state can explain the observed thermovoltage signal, we have calculate the electric, thermoelectric and thermal transport coefficients within the Laudauer formalism of resonant tunneling \cite{Guttman:PRB:51:17758:1995}. The total generalized current through the quantum dot is given by
\begin{equation}
J_{tot}=\int^\infty_{-\infty}dE\left(\frac{\Lambda}{h}\right)[f_L(E,T)-f_R(E,T)]t(E)\label{eqnjtotal}
\end{equation}
where $f_L$ $(f_R)$ is the Fermi distribution function of the left (right) contact, $t(E)$ is the energy dependent transmission coefficient of the QD, and $\Lambda$ is either the charge $(-e)$ or the energy $(E-\mu)$ which is transported through the QD, as applicable to for charge or heat currents, respectively. The charge ($I$) and heat ($Q$) currents are related to the applied electrochemical potential and temperature difference via the transport coefficients $L_{ij}$:
\begin{equation}
\left( \begin{array}{c} I \\
 Q
\end{array}\right)
= \left( \begin{array}{cc} L_{11} & L_{12} \\
 L_{21} & L_{22}
\end{array}\right)
\left( \begin{array}{c} \mu_L/e-\mu_R/e\\
 T_L-T_R
\end{array}\right)\,, \label{eqntranscoeff}
\end{equation}
where $L_{11}=G$ and the thermopower is defined by the quotient $S=-({L_{12}}/{L_{11}})$.

\begin{figure}
\centering
\includegraphics[width = 9.5 cm]{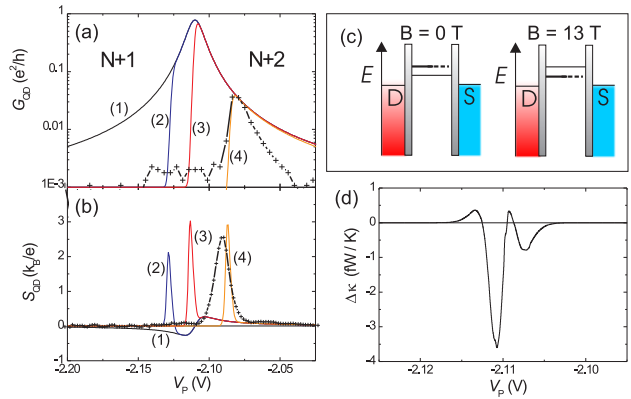}
\caption{(color online) Calculated conductance $G$ (a) and thermopower $S$ (b) for $T_e = 80$~mK, $\Delta T = 30$~mK and various energy separations between the symmetric and asymmetric (blocking) states: (1) without blocking state, (2) $\delta E_Z=-250$~$\mu$eV, (3) 0~$\mu$eV, and (4) 400~$\mu$eV. The crosses represent the measurement at $B_{||}=13$~T. (c) presents a schematic energy diagram of the QD close to the $N+1 \leftrightarrow N+2$ SET resonance at $B_{\parallel}=0$~T and 13~T. The blocking state is symbolized by the thick broken line. (d) displays the difference of the thermal conductances $\Delta \kappa$ for a reversal of the temperature difference.}\label{fig4}
\end{figure}

The energy dependent transmission coefficient can be modeled as  follows: 
\begin{equation}
t(E,T)=A \frac{(\Gamma/2)^2}{(\Gamma/2)^2+ (E)^2}\times f(E-\delta
E_Z,T)\mbox{,}\label{eqntransmissfctn}
\end{equation}
The first term of Eq.\ \ref{eqntransmissfctn} represents a single QD resonance, while the second term accounts for the (thermal) occupation of the blocking state which depends on the energy separation $\delta E_Z$ between the transmitting and the blocking state. 

Figure \ref{fig4} (a) shows the calculated conductance for various energy separations. The values of $A=0.79$ and $\Gamma=0.2$~meV result from a fit of resonance B for $B_\parallel=0$~T. A constant background $G_{\rm cot}=0.001\text{ }e^2/h$ has been added in order to account for co-tunneling contributions via additional excited states. The amplitude of $G_{\rm cot}$ corresponds to the experimentally observed minimum conductance between two SET conductance peaks. Furthermore, the experimentally determined electron temperature, $T_e=80$~mK, and temperature difference of $\Delta T=30$~mK have been used for all calculations.

Starting from a situation where the blocking state is energetically above the symmetric state ($\delta E_Z = -250$~$\mu$eV) the SET conductance peak becomes more and more asymmetric with increasing $\delta E$ (c.f.\ Fig.\ 4 (a)). For $\delta E_Z = 0$ a sharp suppression of the conductance at the center of the SET peak is observed. The calculated conductance for $\delta E_Z = 400$~$\mu$eV [curve 4 in Fig.\ 4 (a)] closely resembles the measured $G(V_{\rm P})$ at $B_{||}=13$~T. This value of $\delta E_Z$ corresponds to a situation where the blocking state is energetically below the symmetric state [Fig.~\ref{fig4} (c)]. The corresponding thermopower [Fig.~\ref{fig4}(b)] changes from a resonance-like line shape to a single peak structure, where good agreement with the experiment (dashed line) is again achieved for $\delta E_Z = 400$~$\mu$eV. 

The change of the line shape for the thermopower is direct evidence that the QD acts as a thermal rectifier. Fig.~\ref{fig4}(d) shows the difference between the thermal conductance, $\Delta \kappa = \kappa_{T_1T_2} - \kappa_{T_2T_1}$, with $\kappa = L_{22} - (L_{12} L_{21} / L_{11})$ and the $L_{i,j}$ obtained from the fits in Figs.\ 4 (a) and (b), for a reversal of the temperature difference across the QD and $\delta E_Z = 400$~$\mu$eV.  At $V_{P} = -2.11$~V the efficiency of rectification, $(\Delta \kappa / \overline{\kappa})$ reaches $10.5\%$ at the given temperatures. This number, while modest, is significantly higher than the rectification achieved in Ref.\ \cite{Chang:Science:314:1121:2006}, and can be even further  increased by an optimized design of the QD layout.


In summary, we have demonstrated that in a QD asymmetric coupling to its leads not only modifies the charge transport, but also strongly modifies its thermal properties. If the QD design is such that transport is favorable through states with non-zero orbital momentum, high effective thermal rectifying properties can be obtained. In contrast to the previously reported nanotube based thermal rectifiers \cite{Chang:Science:314:1121:2006}, QDs directly control the heat transfer in the electronic system, without the need for additional coupling to the phonon system. This opens the perspective of more sophisticated electronic devices with high rectifying performances.

\acknowledgments

We gratefully acknowledge the financial support of the Office of Naval Research (04PR03936-00).

\end{document}